\documentclass[preprint]{aastex}

\usepackage{graphicx,times}             
\usepackage{subfigure}
\usepackage{natbib}
\usepackage{textcomp}
\usepackage{mathtext}
\usepackage{amssymb,amsmath}
\usepackage{longtable,booktabs}

\newcommand{\fermi}{\textit{Fermi}}
\newcommand{\gr}{$\gamma$-ray}

\defcitealias{dai+16}{Paper I}

\begin{document}

\title{Identification of Candidate Millisecond Pulsars from \textit{Fermi} LAT Observations II}

\author{Xuejie Dai\altaffilmark{1,2}, Zhongxiang Wang\altaffilmark{1}, V. Jithesh\altaffilmark{1}
      and Yi Xing\altaffilmark{1}
      }
      
\altaffiltext{1}{Shanghai Astronomical Observatory, Chinese Academy of Sciences,
             Shanghai 200030, China; {\it wangzx@shao.ac.cn}}
\altaffiltext{2}{Graduate University of the Chinese Academy of Sciences, No. 19A, Yuquan Road, Beijing 100049, China}

\begin{abstract}
Following our work presented in Dai et al. (2016), we report our 
detailed data analysis for another 38 \textit{Fermi} $\gamma$-ray 
un-associated sources. These sources are selected from the \textit{Fermi}
Large Area Telescope (LAT) third source catalog on the basis of the properties
of known $\gamma$-ray millisecond pulsars (MSPs) and for the purpose of
finding likely candidate MSPs. From our analysis of the LAT data, 
we identify that among the 38 sources, 28 of them are single point-like 
sources with clean background and their spectra show significant curvature. 
We also conduct analysis of archival X-ray data available for 24 of 
the 28 sources. In the fields of 10 sources, there are at least one X-ray 
object, and in those of the other 14 sources, no X-ray object is detected 
but probably due to the X-ray observations being short. We discuss 
the possible MSP nature for these sources. Six of them
(J0514.6$-$4406, J1035.7$-$6720, J1624.2$-$4041, J1744.1$-$7619, 
J1946.4$-$5403, and J2039.6$-$5618) are most likely 
associated with pulsars because of multi-wavelength identifications 
including direct radio or $\gamma$-ray detection of pulsations. 
To firmly establish the associations or verify the MSP nature for other 
sources, deep X-ray and/or optical observations are needed.
\end{abstract}


\section{Introduction}
\label{sect:intro}

The launch of the \textit{Fermi Gamma-Ray Space Telescope (Fermi)} in
2008 June marks a new era in \gr\ astronomy.  With its unprecedented
capabilities, the Large Area Telescope (LAT) onboard \fermi\ has allowed us 
to, for the first time, find large numbers of different classes 
of \gr\ sources in the sky and study their properties in detail.
Using the first four-years (year 2008--2012) all-sky monitoring data obtained 
with LAT, more than 3000 sources in the energy range of 0.1--100 GeV
have been found \citep{3fgl15}.  Classification studies of these sources,  
those contained in the \fermi\ LAT third source catalog (3FGL),
have confirmed the results from the surveys of the sky with previous 
Gamma-Ray telescopes, e.g., the \textit{Compton Gamma-Ray Observatory}: 
the dominant class of the \gr\ sources is Active Galactic Nuclei 
(AGN; \citealt{3fagn15}) and in the Milky Way, pulsars are the 
majority \citep{2fpsr13,3fgl15}.

From \fermi\ LAT observations of pulsars, it has been learned that
they have stable \gr\ emission and their spectra can generally be described
by a power law (PL) with an exponential cutoff \citep{2fpsr13}.
Such spectral properties well match the theoretical expectations
for the high-energy emission mechanisms of pulsars (e.g., \citealt{mh04}),
and can be used for finding candidate new pulsars among the nearly
1000 3FGL sources that have not been found to be associated with any 
known types of high-energy objects \citep{3fgl15}. Currently more 
than 200 \gr\ pulsars have been 
identified\footnote{https://confluence.slac.stanford.edu/display/GLAMCOG/Public+List+of+LAT-Detected+Gamma-Ray+Pulsars}, 
and among them more than 20 are newly discovered millisecond pulsars (MSPs). 
The discoveries of the significant number of new MSPs were made due to
\fermi\ LAT's first detection of them, so that follow-up radio and other 
wavelength observations could be carried out for identification.

For the purpose of finding new MSPs, we have conducted a systematic
study of un-associated 3FGL sources. We have selected 101 sources from
3FGL, requiring the properties of being non-variables with curved
spectra and having Galactic latitudes of $>$5 degrees 
(\citealt{dai+16}; hereafter paper I). There were 24 sources with a low 
detection significance (average\_sig$<$6; \citealt{3fgl15}).
No data analysis was conducted for the 24 sources because of their low
detection significance: the low photon counts do not allow us to clearly 
determine their properties.  For the remaining 77 sources, their \fermi\ LAT
data were analyzed. From the analysis, those contaminated
by extended background emission or mixed with nearby unknown sources were
excluded. The further selection was conducted on the basis of the spectra
we obtained. In this way, we were able to find `good' candidate MSPs
for follow-up identification. In \citetalias{dai+16},
we reported our target selection, and because large amount of computing
time is required for \fermi\ data analysis,  the detailed LAT
data analysis for 39 of them was presented. The sources are mostly in 
the Northern Hemisphere, and we were able to
find 24 from them as possible candidate MSPs (which
were thus taken as the targets for our follow-up observation program 
conducted with optical telescopes).
In this paper, we report our
data analysis for the other 38 sources that are in the Southern Hemisphere.

\section{\fermi\ LAT Data Analysis}
\label{sec:ar}

\subsection{\fermi\ LAT Data}

LAT onboard \fermi\ is an instrument generally carrying out
an all-sky survey in the energy range from 20 MeV to 300 GeV. 
With its wide field-of-view and high sensitivity, 
$\gamma$-ray events are distinguished from   
background events through measuring the direction, energy, and arrival 
time of each $\gamma$-ray photon \citep{atw+09}.
In our data analysis, we used the latest Pass 8 data, which
were from 2008 August 4 15:43:39 to 2015 October 22 00:26:36 (UTC).
We extracted data within 15 deg of a target's position in the energy 
range from 200 MeV to 300 GeV, for which
photons below 200 MeV were not included to avoid the 
relative large uncertainties of the instrument response function of 
the LAT in the low energy range. In addition, as recommended by the LAT team, 
we selected events with zenith angles 
less than 90 deg to exclude possible contamination from the Earth's limb. 

\subsection{Maximum Likelihood Analysis}

Using the newly released LAT science tools package v10r0p5, we performed 
a standard binned maximum likelihood analysis \citep{mat+96} to the data 
of each target.  In a source model for a target, all sources within 
the 20 deg region were included. The spectral parameters of these sources 
are provided in 3FGL, and the spectral normalization parameters of 
those within 5 deg from each target were set free and all the other parameters 
were fixed at their catalog values. For the Galactic 
and the extragalactic diffuse emission, we included 
the model gll\_iem\_v06.fits and spectrum file 
iso\_P8R2\_SOURCE\_V6\_v06.txt in the source model. 
The normalization parameters of the two diffuse emission components 
were left free.

We obtained the Test Statistic (TS) map of a 2$^\circ\times 2^\circ$ 
region centered at the position of each target. 
TS values are calculated from TS$=-2\log (L_{0}/L_{1})$, 
where $ L_{0} $ and $ L_{1} $ are the maximum likelihood values
for a model without and with an additional source at a specified 
location respectively \citep{1fgl10}. The TS value 
for a given source is approximately the square of 
the detection significance.  We examined the TS map of each target, 
and selected `clean' sources among the targets, which we defined to
be point-like sources, not mixed with other unknown 
sources and/or not in a region with strong extended 
emission (see examples in \citetalias{dai+16}). We were able to find 29 sources 
as such clean sources. They are listed in 
Tables~\ref{tab:cmsps} \& \ref{tab:nosc}.
We then ran \textit{gtfindsrc} in the LAT software package to 
determine the positions for these 29 sources. 
The best-fit positions we obtained are consistent with
those provided in 3FGL within $ 2\sigma $ error circles.


The other 9 sources, which were found not to be clean point-like sources,
were excluded from our target list. For them, 
further data analysis to determine their properties would require 
large amount of computing time.  Their spectral parameters in 3FGL are 
provided here in Table~\ref{tab:bgs}. 
Among them, six sources were fitted with a power law (PL) model, 
\begin{equation}
\frac{dN}{dE} = N_0 E^{-\Gamma}\ \ \ ,
\end{equation}
where $N_{0}$ and $\Gamma$ are the normalization and photon index, 
respectively. 
The other three sources were fitted with a LogParabola model,
\begin{equation}
\frac{dN}{dE} = N_0\left(\frac{E}{E_b}\right)^{-\alpha-\beta log \left(E/E_b\right)}\ \ \ ,
\end{equation}
where $N_{0}$ , $ \alpha $, and $ \beta $ are flux density, photon index, 
and the curvature, respectively. The energy $ E_b $ was set such that errors
on differential fluxes were minimal, and “Signif\_curve” 
(in Table~\ref{tab:bgs}) is the curvature significance estimated from 
likelihood values for a PL model or a LogParabola model.

\subsection{Spectral Analysis}
\label{subsec:sa}

\begin{figure}
	\centering
	\includegraphics[width=0.36\textwidth]{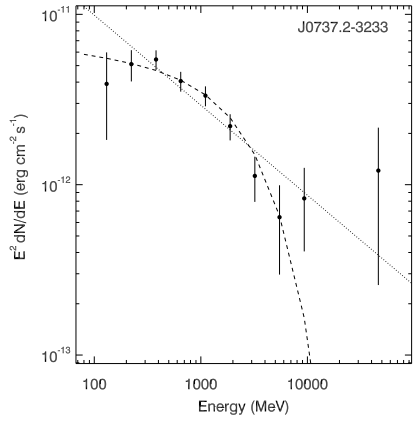}
	\caption{$\gamma$-ray spectrum of J0737-3233. The solid and
dashed lines are the best-fit PL and PLE models, respectively.}
	\label{fig:nmsps}
\end{figure}

To obtain the $\gamma$-ray spectrum for each clean point-like source,
we ran \textit{gtlike} at the best-fit position determined.
We evenly divided energy logarithmically from 0.1 to 
300 GeV into 15 energy bands.
We first modeled each source with a simple PL, with
$\Gamma$ fixed at the value derived from the above.
The spectral data points were obtained, but only those with
TS$>4$ were kept.  We secondly repeated the analysis using a PL 
with an exponential cutoff (PLE),
\begin{equation}
\frac{dN}{dE} = N_0 \left(\frac{E}{E_0}\right)^{-\Gamma}\exp(-\frac{E}{E_c})\ \ ,
\end{equation}
where $E_{c}$ is the cutoff energy 
and $E_0 =1$ GeV was set.
The results from the two spectral
models were compared, and the curvature significance Signif\_curve was 
estimated from Signif\_curve$=\sqrt{2log(L_{PLE}/L_{PL})}$,
where $ L_{PLE} $ and $ L_{PL} $ are the maximum likelihood values modeled
with PLE and PL, respectively.  The analysis shows that all the sources 
had significant curvature except one, J0737.2$-$3233.
Its spectral results are given in Table~\ref{tab:nosc},
indicating that a PLE model is not significantly better than a PL one. 
In Figure~\ref{fig:nmsps}, its spectrum is also shown.
We excluded this source from our candidate list.

\subsection{Variability Analysis}

We performed temporal analysis of the LAT data for 
the 28 remaining sources. The time period was from 2008 August 4 
23:59:59 to 2015 September 30 23:59:56(UTC) and we divided it into 
30-day intervals.  The PL model, with photon index fixed at the value 
obtained in Section~\ref{subsec:sa}, was used for conducting likelihood 
analysis in each time bin at the best-fit position of each source. 
We obtained the light curves and TS curves for the sources.
Consistent with the results in the LAT third source catalog,
we did not find any significant flux variations in the light curves.

\section{X-ray Data Analysis}  

The possible X-ray counterparts of the 28 candidate MSPs were 
searched in archival X-ray observations. We utilized publicly 
available X-ray 
observations from {\it Swift}, {\it Chandra}, {\it XMM-Newton}, 
and {\it Suzaku} 
satellites. Among the candidates, 26 of them were observed at-least once 
with the above mentioned observatories, and we selected the longest-exposure 
observation among the available datasets. However, two sources 
\citep[J1744.1$-$7619 and J2039.6$-$5618;][]{hui+15,sal+15,rom15} 
have been well studied in multi-wavelength and verified to likely be MSPs. 
These two sources were thus not included in this analysis. We used the HEASOFT 
package version 6.15.1 distributed by the High Energy Astrophysics Science 
Archive Research Center (HEASARC) for the analysis. 

\subsection{Data analysis}
\subsubsection{{\it Swift} XRT Analysis}
\label{subsec:swift}

The data for the candidate MSPs obtained with the {\it Swift} X-ray 
Telescope \citep[XRT;][]{bur+05} were downloaded from HEASARC. 
The unfiltered event files from Photon Counting (PC) 
mode observations were reduced using the {\sc xrtpipeline} task and 
calibration files version 20150721 available in the {\it Swift} CALDB. 
The X-ray sources in the XRT images were detected from running the command 
{\sc detect} available in the {\sc ximage}. 
We used a detection threshold of $3\sigma$ 
and considered only the X-ray sources within the $2\sigma$ {\it Fermi} error 
circle as the possible X-ray counterparts. The exact positions of the X-ray 
sources detected were estimated with using the {\sc xrtcentroid} task. 
For the possible counterparts, we extracted the source and background 
spectra from a circular region of radius 47 arcseconds along with 
the ancillary response files (ARF) and response matrix files (RMF). 
If there were sufficient spectral counts to perform spectral 
modeling, we grouped the spectra using {\sc grppha} with a minimum of 
20 counts per bin and adopted the $\chi^2$ statistic. For the sources with 
limited net counts, we used the Cash Statistic \citep{cash79} for 
the spectral fitting. 

For the detected X-ray sources, we tested the spectral models such as PL, 
blackbody ({\tt BBODY}), and {\tt APEC}, each combined with 
interstellar absorption. 
In all cases, we fixed the absorption column 
density \citep[{\tt tbabs};][]{wil+00} 
to the Galactic values \citep{kal+05}. The spectral fitting results with 
an absorbed PL model are summarized in Table \ref{xray}. In many cases, 
no X-ray source was detected, thus we estimated the $3\sigma$ upper limits
on the count rates using the {\sc uplimit} command in {\sc ximage}. 
The upper limits were then converted into fluxes with 
the webPIMMS\footnote{http://heasarc.gsfc.nasa.gov/cgi-bin/Tools/w3pimms/w3pimms.pl} used (assuming an absorbed PL model with $\Gamma= 1.7$ and 
the absorption column density of the Galactic value).
The upper limits are listed in the Table \ref{nodetect}.     

\subsubsection{{\it Chandra} Analysis}

The {\it Chandra} observations were analyzed using 
the science threads of {\it Chandra} Interactive Analysis of Observations
(CIAO) version 4.6 with CALDB version 4.6.1.1. The {\it Chandra} data were 
reprocessed with the CIAO tool {\sc chandra\_repro}. We ran 
the {\sc celldetect} task on the reprocessed event files with a detection 
threshold of $3\sigma$ to detect the X-ray sources. The source and background 
regions were extracted from a circular regions of radius 5--10 arcseconds,
and we performed the {\sc specextract} task in CIAO to generate the source 
and background spectra and the corresponding response files (ARF and RMF). 
The source spectra were grouped using {\sc grppha} with minimum counts 20 
per bin, and the spectral fitting was performed with the $\chi^2$ statistic. 
For the low quality spectra, we used the Cash Statistic.     

\subsubsection{{\it XMM-Newton} Analysis}

We retrieved the observation data files from the {\it XMM-Newton} Science 
Archive and used the {\it XMM-Newton} Science Analysis Software (SAS) 
version 14.0 to analyze them.  We performed standard data processing for 
the European Photon Imaging Camera (EPIC) pn \citep{str+01} and 
MOS \citep{tur+01} detectors with the {\sc epchain} and {\sc emchain} tools. 
The high particle background time intervals were excluded from 
the observations and only 0--4 pattern events from 
the pn and 0--12 from the two MOS detectors were selected. 
We ran the detection on the cleaned and filtered event files  
in the 0.3--10 keV energy range using the SAS task {\sc edetect\_chain}. 
As mentioned in the Section \ref{subsec:swift}, 
we selected the X-ray sources within the $2\sigma$ {\it Fermi} error circles
and extracted the source and background spectra from circular regions of 
radius 12--30 arcseconds. The source and 
background spectra, together with response and ancillary response files, 
were obtained using the {\sc especget} task. For each source, 
we fitted simultaneously the pn and MOS spectra 
using XSPEC version 12.8.1g. The spectral modeling was performed with
either the $\chi^2$ statistic or the Cash Statistic. The resulting
spectral model parameters are given in Table \ref{xray}.    

\subsubsection{{\it Suzaku} Analysis}

\begin{figure}
	\centering
	\includegraphics[width=0.4\textwidth, angle=0]{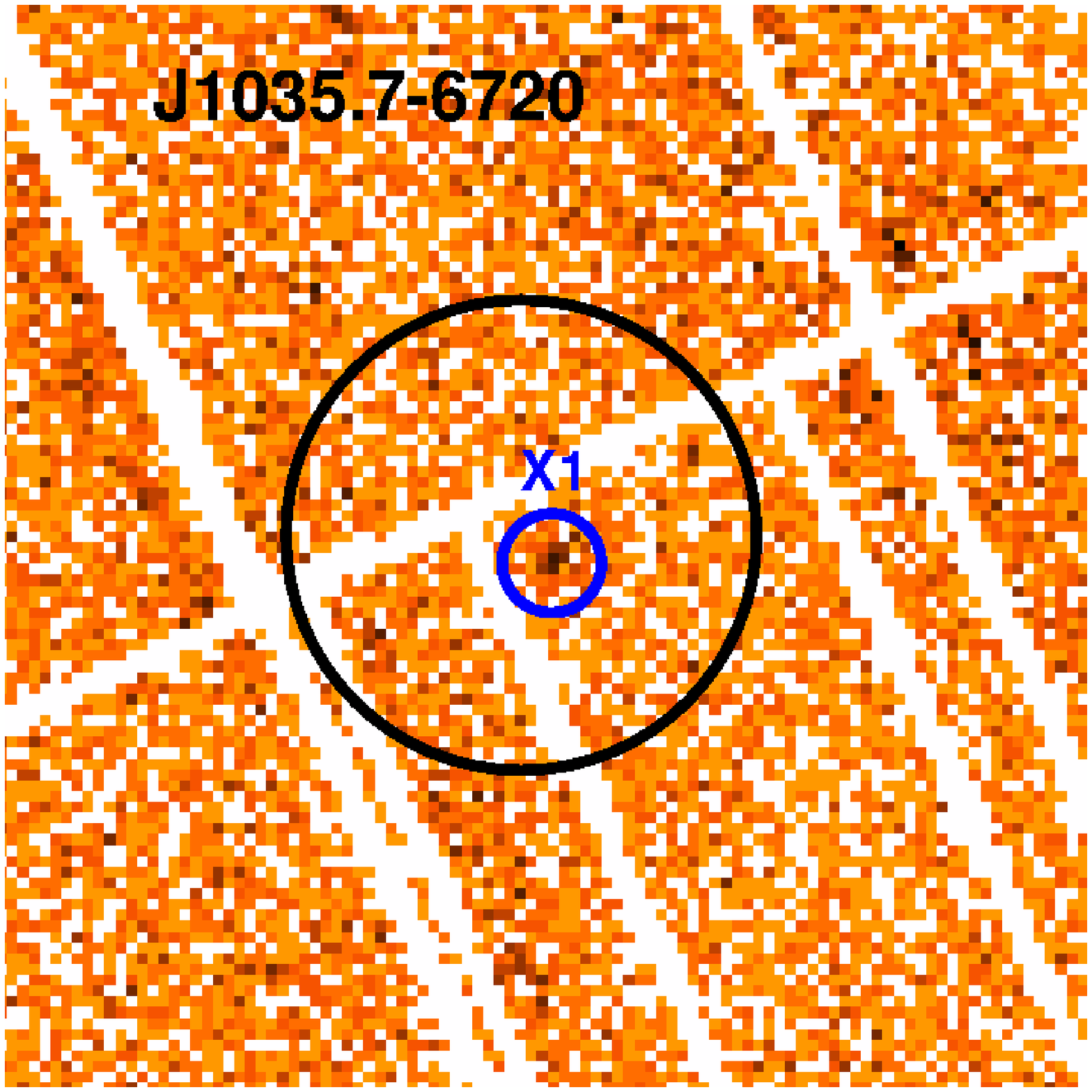}
	\includegraphics[width=0.4\textwidth, angle=0]{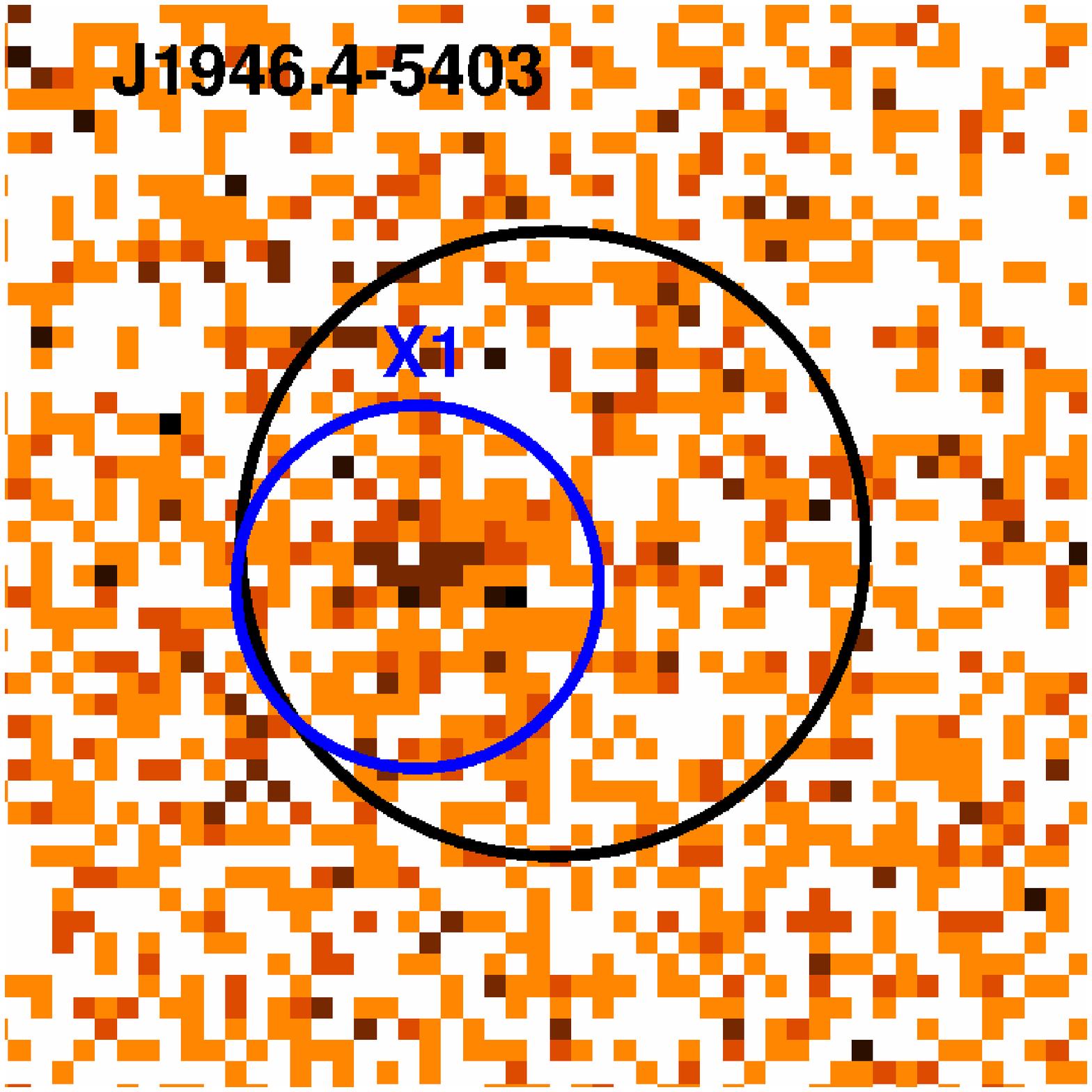}
	\caption{Image fields of J1035.7$-$6720 ({\it left}) and 
J1946.4$-$5403 ({\it right}) as obtained from the {\it XMM-Newton} EPIC-PN 
and {\it Suzaku} XIS, respectively (see Table~\ref{xray}). 
The black circles indicate the $2\sigma$
{\it Fermi} error circles and the candidate X-ray counterparts are marked by 
blue circles. }
\label{fig:xray}
\end{figure}

{\it Suzaku} \citep{mit+07} observed the candidate J1946.4-5403 
with its X-ray imaging spectrometer (XIS) on 2011 October 31 for 
an exposure time of 42.4 ks (Observation ID: 706026010). We cleaned and 
calibrated the unfiltered event files (XIS data) 
using standard filtering criteria with the specific {\sc headas} 
tool {\sc aepipeline} and calibration files (version 20130110) available in 
the {\it Suzaku} CALDB. The source, shown in Figure~\ref{fig:xray},
and background regions 
were taken from a circular region of radius $70 \arcsec$ and 
the spectra, response matrices, and ancillary response files were generated
using {\sc xselect} for XIS0, XIS1 and XIS3. The front illuminated (FI) CCDs 
spectra, XIS0 and XIS3, were added using the {\sc ftool addascaspec}. 
The co-added spectrum was then grouped 
to minimum counts of 20 and $\chi^2$ statistics was used for 
the spectral fitting.

\subsection{Individual Sources}
Among the 24 sources studied here, 10 of them have one or more X-ray 
sources within the $2\sigma$ \fermi\ error circle and the 
rest of them do not have.
In the following sections, we discuss the properties of 
individual sources which we fitted with spectral models other than a PL.

\subsubsection{J0838.8$-$2829}

We have detected only one source, named as X1, 
for J0838.8$-$2829 in the {\it Swift} observation (ObsID: 00041343002). 
We examined the spectrum with an absorbed PL model and the
spectral results are given in Table \ref{xray}. We then added a blackbody 
component ({\tt BBODY}) to the PL and the spectral fit was improved 
marginally ($\Delta\chi^2 \sim$ 3.7 for 2 degrees of freedom [dof] 
at a confidence level of $\sim 82\%$). The spectral parameters,
$kT= 0.14^{+0.06}_{-0.09}$ keV, $\Gamma =1.46^{+0.26}_{-0.30}$, and
$F_{X} = 5.78^{+4.42}_{-4.40}\times~10^{-13}~\rm erg~cm^{-2}~s^{-1}$
($\chi^2/\rm dof$ =11.4/12), are consistent with those of typical 
MSPs \citep{zav07, mar12}.  

\subsubsection{J0933.9$-$6232}

The 43 ks {\it Chandra} ACIS-S observation detected only one X-ray 
source (X1) in the field of J0933.9$-$6232. The absorbed PL model did not 
provide a statistically acceptable fit ($\chi^2/\rm dof$ =25.9/7) 
for the spectrum of this source. 
We also tried different models such as a {\tt BBODY} or an {\tt APEC}.
In all cases, the spectral fits were worse ($\chi^2_{r} > 2$). Thus, we 
used two-component model such as a PL plus {\tt APEC} model. This model 
gave an acceptable fit, $\chi^2/\rm dof$ =8.6/5, 
from which $\Gamma =2.22^{+1.99}_{-2.87}$, $kT= 0.90^{+0.11}_{-0.14}$ keV, 
and $F_{X} = 1.54^{+3.84}_{-0.68}\times~10^{-14}~\rm erg~cm^{-2}~s^{-1}$.  
We note that \citet{saz+16} also analyzed the same observation, and reported
the detection of X1. However, only the PL model was considered in their
analysis.

\subsubsection{J1119.9$-$2204} 

The field of J1119.9$-$2204 was observed by {\it Swift} on multiple 
occasions (ObsID: 00049351025 and 00049351021) and \citet{hui+15} have 
reported their data analysis. Two sources were detected with {\it Swift}.
{\it XMM-Newton} observed the field
on 2014 June 14 for a total exposure of $\sim 73.6$ ks (Obs. ID: 0742930101). 
In the {\it XMM-Newton} observation, one additional source was detected.
Among them, the source X1 
had enough spectral counts, allowing detailed spectral modeling. We 
initially fitted the spectrum with 
an absorbed PL model which provides an acceptable fit,
with $\Gamma =2.21\pm0.15$ ($\chi^2/\rm dof$ =45.8/33). The addition 
of a blackbody component to the PL improved the 
spectral fit significantly, $\Delta\chi^2 \sim$ 12.5 for 2 dof at 
a confidence level of $> 99\%$. 
The best-fit parameters are $kT= 0.20\pm0.03$ keV, 
$\Gamma =1.34^{+0.62}_{-0.78}$, and 
$F_{X} = 5.42^{+1.65}_{-2.50}\times~10^{-14}~\rm erg~cm^{-2}~s^{-1}$.
The parameters of X1 were consistent with those of 
typical MSPs \citep{zav07,mar12}, while the emission from each of the
other two sources (X2 and X3) favors a non-thermal case  (Table~\ref{xray}).   

\subsubsection{J1624.2$-$4041}

The field of J1624.2-4041 was observed by {\it XMM-Newton} for 
an exposure of 31 ks on 2013 August 14 (ObsID: 0722940101). 
We identified 8 possible X-ray sources 
and analyzed the spectra of these sources with an absorbed PL model. 
All the sources were well fitted except the source X6 (See Table \ref{xray}). 
For X6, the PL model could provide a statistically acceptable fit 
(C = 8.2 for 8 dof), but resulted in 
an unphysically large photon index, $\Gamma =6.18^{+3.15}_{-2.80}$. 
Therefore, the non-thermal case is not favored for this source. We 
tested an absorbed blackbody model. With $\rm N_{H}$ 
fixed at the Galactic value $2.60\times 10^{21}~\rm cm^{-2}$, the model
yielded a temperature of $kT= 0.12^{+0.06}_{-0.03}$ keV with 
C = 8.8 for 8 dof, which indicates a possible thermal origin for this source.  

\subsubsection{J1626.2$-$2428c}
\label{sec:j1626}

The deep {\it Chandra} ACIS-I observations 
(ObsID: 17249 and 637; Exposure times of 100.1 and 
97.7 ks respectively) conducted on 2014 October 06 and 2000 May 15 covered 
the field of J1626.2$-$2428c. We reprocessed these long observations 
using standard tools 
in CIAO and identified two X-ray sources. The emission from the 
first source (X1) can be fitted by a PL model in both observations. 
The PL index and flux are constant within the uncertainties 
(see Table \ref{xray}). The spectra of the source X2 cannot be described 
by a single component model (PL, blackbody, or {\tt APEC}) or
two-component models (such as a PL plus blackbody 
or a PL plus {\tt APEC}). Thus, we attempted a broken PL model 
in both observations. 
For the 2014 October observation, the broken PL fit results 
in $\Gamma_{1} = -1.85^{+0.87}_{-1.13}$ 
below the break energy $E_{\rm break} = 3.08^{+0.27}_{-0.21}$ and $\Gamma_{2} = 2.22^{+0.66}_{-0.55}$ 
above the break energy with $\chi^2/\rm dof$ = 19.2/15. 
For the 2000 May observation, the broken PL provided an acceptable fit 
with $\Gamma_{1} <-0.31 $, 
$E_{\rm break} = 2.70^{+0.31}_{-0.93}$, and
$\Gamma_{2} = 1.87^{+0.58}_{-1.42}$ ($\chi^2/\rm dof$ = 5.3/7).
These spectral parameters have been widely reported for 
high-redshift radio loud quasars \citep{fab+01, yuan+06, do16}. 
Thus from the X-ray properties we can say that X2 is possibly 
a background quasar.

\section{Results and Discussion}
\label{sec:rd}

\begin{figure}
   \centering
   \includegraphics[width=0.36\textwidth]{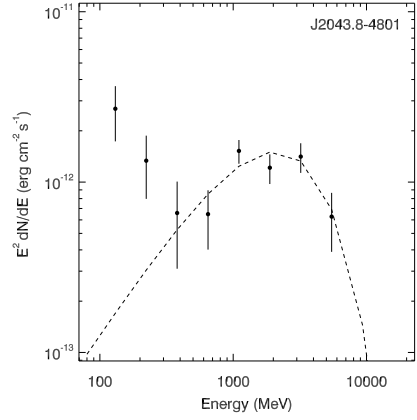}
   \caption{$\gamma$-ray spectrum of J2043.8-4801.
The dashed curve indicates the best-fit PLE model, which does not well
describe the spectral data points as the spectrum probably has two components.
    }
   \label{fig:tcs}
\end{figure}

We have analyzed the \fermi\ LAT data for another 38 un-associated sources
selected from 3FGL for identifying candidate MSPs.
From the analysis, 29 clean point-like sources were found, 
while one of them, J0737.3$-$3233, was excluded as a candidate MSP 
since its spectrum does not show significant curvature. 
Among the remaining 28 sources, J2043.8$-$4801 appears to possibly
have two components in its spectrum, which is shown in Figure~\ref{fig:tcs}.
We examined its TS maps and it is consistent with being a point source.
The TS maps at low and high energy ranges 
(e.g., 0.1--1.0 GeV and 1.0--300\,GeV, respectively) 
were also calculated, but no evidence was found for 
having an additional source at the source position.
Such a source needs further studies for investigating the possible nature 
of its emission.

Five of the 28 sources have already been identified
as likely pulsars from radio searches or $\gamma$-ray data analysis: 
J0514.6$-$4406 is probably the counterpart to
PSR J0514$-$4407 (but with a spin period of 302.2 ms; \citealt{bha+16}); 
J1946.4$-$5403 is 
an MSP in a 3-hours binary (a possible black widow system; 
\citealt{cam+15}); J1035.7$-$6720, J1624.2$-$4041, and J1744.1$-$7619
have been found with $\gamma$-ray pulsations \citep{cam+15}.
Optical and/or X-ray observations have also helped identification of
possible MSP nature for several sources:
J2039.6$-$5618 is very likely a redback
MSP system with an orbital period of $\sim$5.4 hours \citep{sal+15,rom15};
J0933.9$-$6232 (\citealt{saz+16} and this work) and 
J1119.9$-$2204 (2FGL J1120.0$-$2204; \citealt{gui+12,hui+15}; this work)
were found with possible X-ray counterparts and suggested to likely be MSPs.
These sources are marked as `c-MSP' in Table~\ref{tab:cmsps}.
In addition we note in \citet{cam+15}, 13 of our candidates are listed as 
their targets for searching for radio pulsars, although thus far no radio
pulsation signals have been found yet.
These studies strongly support our data analysis results.

\begin{figure}
   \centering
   \includegraphics[width=0.36\textwidth]{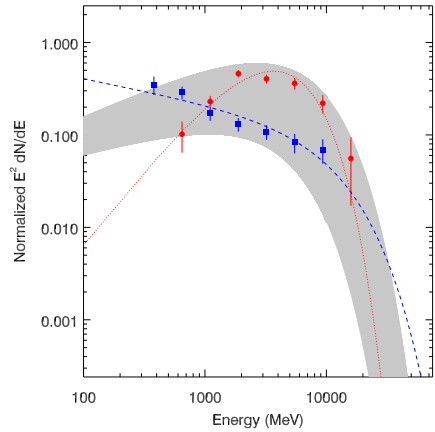}
   \caption{$\gamma$-ray spectra of J1539.2$-$3324 (red dots) and 
J1626.2$-$2428c (blue squares), with the red dotted and blue dashed
curves being the best-fit PLE models, respectively.
The grey area indicates the 3$\sigma$ region of the best-fit spectral model
derived from 39 known $\gamma$-ray MSPs listed in \citet{2fpsr13}.
    }
   \label{fig:model}
\end{figure}

MSPs generally have \gr\ spectra with $\Gamma$ and $E_c$ in the ranges 
of 0.4--2.0 and 1.1--5.4\,GeV respectively (see \citealt{2fpsr13} for details).
Recently, \citet{xw16} have analyzed the spectra of 39 MSPs listed in the second
\fermi\ \gr\ pulsar catalog and obtained the spectral ranges of 
$\Gamma=1.43$--1.64 and $E_c=3.00$--4.65\,GeV (3$\sigma$) from fitting
their spectra with a PLE model.
If we consider the $\Gamma$ range, which is relatively well determined,
five sources have their $\Gamma$ values out of the range, especially 
the sources J1539.2$-$3324 and 
J1626.2$-$2428c that have the parameters of $\Gamma=0.4\pm0.2$, 
$E_c=2.2\pm0.3$\,GeV and $\Gamma=2.3\pm0.1$, $E_c=10\pm5$\,GeV, respectively. 
For this reason, we tentatively suggest that the five sources are not
MSPs, marked with `N' in Table~\ref{tab:cmsps}.
The spectra of J1539.2$-$3324 and J1626.2$-$2428c are shown as an example in 
Figure~\ref{fig:model}. The first source 
has been searched
in X-ray data as a candidate MSP but no possible counterpart
was found \citep{hui+15} and is listed as a candidate blazar 
in the \fermi\ third AGN catalog \citep{ack+15}. The second one, with
a larger $\Gamma$ value, is possibly associated with the blazar 
PMN~J1626$-$2426 \citep{ack+15}. However if this is the case, none of 
the two X-ray sources reported in Section~\ref{sec:j1626} would be 
the counterpart because of the position mismatch \citep{gw93}. 
On the other hand,  the other three `N' sources 
include J0933.9$-$6232, which is considered as a promising MSP candidate. 
Whether the $\Gamma$ range derived in \citet{xw16} is a reliable 
criterion for excluding non-MSP sources will be
tested, once the nature of such `N' sources are identified.

We have analyzed archival X-ray data that were available for 24 of our 
selected candidate MSPs. In the fields of 10 of them, at least one object 
was detected. 
For the likely pulsars J1035.7$-$6720, J1624.2$-$4041, and J1946.4$-$5403, 
the spectra of the objects are consistent with being a non-thermal PL with
photon indices of $\sim 1$--3.
In addition, the flux ratios between their $\gamma$-ray
0.1--100 GeV flux ($G_{100}$) and the X-ray fluxes of the detected objects
are in a range of $\sim$100--1000, also consistent with that of the most 
known $\gamma$-ray pulsars \citep{2fpsr13}. Both properties suggest that
the X-ray objects, particularly in the fields of J1035.7$-$6720 and 
J1946.4$-$5403 (containing only one object), are likely the counterparts.
While detailed X-ray studies of the objects in the fields of
J0933.9$-$6232 and J1119.9$-$2204 have been previously conducted, our
analysis of the different sets of X-ray data have confirmed the previous 
results.
In order to identify the X-ray objects in each source field and possibly verify
their pulsar nature, optical observations of them for searching for 
variability can be the next step. Nearly 70\% of the \fermi\ 
$\gamma$-ray MSPs are in a binary (e.g., \citealt{2fpsr13,cam+15}).
Since the companions of MSP binaries are irradiated by pulsar winds,
they may show significant orbital modulation at optical bands, 
thus revealing their pulsar binary nature (e.g., \citealt{rom15}).  
Finally 14 of
our selected candidates were not detected with any X-ray objects in
their fields, the reason for which is probably because 
the X-ray observations were not sufficiently deep.
The lower limits on the $\gamma$-ray--to--X-ray flux ratios 
are only as large as $>$100 (see Table~\ref{nodetect}).
Deep X-ray observations are needed in order for finding X-ray 
objects and thus allowing further multi-wavelength studies.

As a summary for this systematic study, we have selected 101 sources from 
the \fermi\ 3FGL un-associated sources on the basis of the properties
of the known $\gamma$-ray MSPs and conducted detailed LAT data
analysis for 77 of them that were detected with a detection
significance of $\geq 6$. Our analysis indicates that 52 sources
are point-like without strong background contamination and their emission
is well described by a PLE model. A few of the 52 sources have already
been studied at multiwavelengths and shown to be likely pulsars. 
We consider the remaining ones as good candidate
MSPs for follow-up identification studies.
For 44 of these candidates, we have conducted analysis 
of the archival X-ray observations. We have obtained spectral properties
of the X-ray objects detected in the fields of 14 candidates, and derived
flux upper limits for the other 30 candidates that did not have
any X-ray objects detected in their fields (\citetalias{dai+16} and this work).
The X-ray study results
generally support their pulsar nature. Finally we note that approximately
10 of our candidates (excluding those already identified as pulsars) 
are also listed as promising dark matter subhalo 
candidates (marked with c-subhalo in Table~\ref{tab:cmsps}; see also
\citetalias{dai+16}) in \citet{bhl15}. This possibility certainly
makes the candidates as more interesting targets for follow-up studies.


\begin{acknowledgements}
This research made use of the High Performance Computing Resource in the Core
Facility for Advanced Research Computing at Shanghai Astronomical Observatory.
This research was supported by the National Program on Key Research and 
Development Project (Grant No. 2016YFA0400804), the Strategic Priority 
Research Program ``The Emergence of Cosmological Structures" of 
the Chinese Academy of Sciences (Grant No. XDB09000000), and
the National Natural Science Foundation of China (11373055, 11633007). 
Z.W. acknowledges the support 
by the CAS/SAFEA International Partnership Program for Creative Research Teams;
J.V. by the Chinese Academy of Sciences President’s International 
Fellowship Initiative (CAS PIFI, Grant No. 2015PM059); and
Y.X. by the Shanghai Natural Science 
Foundation for Youth (13ZR1464400) and the National Natural Science Foundation
of China for Youth (11403075). 
\end{acknowledgements}

\clearpage
\begin{table}
\begin{minipage}[]{100mm}
  \caption[]{Spectral results for 28 candidate MSPs.\label{tab:cmsps}}
\end{minipage}
\fontsize{9}{9}\selectfont
\centering
  \begin{tabular}{llcccccc}
\hline
Source name    & Spectra model    & Flux/$10^{-9}$    & $\Gamma$         & E$_{c}$          & TS   & Signif\_Curve & Comments \\
\qquad         & \qquad  & (photons cm$^{-2}$\,s$^{-1}$) & \qquad        & (GeV)            &\qquad& ($\sigma$)    &  \\
\hline
J0048.1$-$6343 & PowerLaw         & 1.5 $\pm$ 0.3    & 2.1 $\pm$ 0.1     & \qquad           & 87   & 3.48   & \\
\qquad         & PLSuperExpCutoff & 0.8 $\pm$ 0.3    & 0.7 $\pm$ 0.9     & 2 $\pm$ 2        & 98   & \qquad & \\	
\hline 
J0514.6$-$4406 & PowerLaw         & 5.9 $\pm$ 0.5    & 2.58 $\pm$ 0.07   & \qquad           & 261  & 7.72   & c-MSP  \\
\qquad         & PLSuperExpCutoff & 4.3 $\pm$ 0.6    & 0.5 $\pm$ 0.5     & 0.5 $\pm$ 0.1    & 315  & \qquad &   \\	
\hline
J0802.3$-$5610 & PowerLaw         & 9.0 $\pm$ 0.7    & 2.39 $\pm$ 0.06   & \qquad           & 332  & 7.10   &   \\
\qquad         & PLSuperExpCutoff & 7.3 $\pm$ 0.7    & 1.4 $\pm$ 0.2     & 1.5 $\pm$ 0.4    & 382  & \qquad & \\	
\hline
J0838.8$-$2829 & PowerLaw         & 7.3 $\pm$ 0.6    & 2.58 $\pm$ 0.07   & \qquad           & 482  & 5.11   & \\
\qquad         & PLSuperExpCutoff & 5.8 $\pm$ 0.6    & 1.6 $\pm$ 0.1     & 6 $\pm$ 2        & 506  & \qquad & \\	
\hline 
J0933.9$-$6232 & PowerLaw         & 10.1 $\pm$ 0.6   & 2.03 $\pm$ 0.04   & \qquad           & 842  & 12.72  & c-MSP  \\
\qquad         & PLSuperExpCutoff & 6.0  $\pm$ 0.6   & 0.6 $\pm$ 0.2     & 1.9 $\pm$ 0.3    & 996  &        & N \\	
\hline
J0954.8$-$3948 & PowerLaw         & 15.6 $\pm$ 0.8   & 2.60 $\pm$ 0.05   & \qquad           & 696  & 4.26   &   \\
\qquad         & PLSuperExpCutoff & 14.7 $\pm$ 0.9   & 2.2 $\pm$ 0.1     & 4 $\pm$ 1        & 712  &        & N \\	
\hline
J1035.7$-$6720 & PowerLaw         & 18.7 $\pm$ 0.3   & 2.19 $\pm$ 0.01   & \qquad           & 1493 & 12.84  & c-MSP  \\
\qquad         & PLSuperExpCutoff & 14.4 $\pm$ 0.8   & 1.3 $\pm$ 0.1     & 2.2 $\pm$ 0.3    & 1637 & \qquad &  \\	
\hline
J1119.9$-$2204 & PowerLaw         & 17.2 $\pm$ 0.7   & 2.24 $\pm$ 0.03   & \qquad           & 2114 & 12.57  & c-MSP c-subhalo  \\
\qquad         & PLSuperExpCutoff & 13.7 $\pm$ 0.7   & 1.3 $\pm$ 0.1     & 1.7 $\pm$ 0.2    & 2262 & \qquad &  \\	
\hline
J1231.6$-$5113 & PowerLaw         & 13.8 $\pm$ 0.9   & 2.70 $\pm$ 0.06   & \qquad           & 461  & 6.25   &   \\
\qquad         & PLSuperExpCutoff & 12.0 $\pm$ 0.9   & 1.7 $\pm$ 0.2     & 1.1 $\pm$ 0.3    & 492  &        &  \\	
\hline
J1400.2$-$2413 & PowerLaw         & 3.4 $\pm$ 0.5    & 2.15 $\pm$ 0.08   & \qquad           & 169  & 5.95   & \\
\qquad         & PLSuperExpCutoff & 1.8 $\pm$ 0.2    & 0.7 $\pm$ 0.4     & 1.7 $\pm$ 0.5    & 204  & \qquad &\\	
\hline
J1539.2$-$3324 & PowerLaw         & 6.5 $\pm$ 0.5    & 1.89 $\pm$ 0.04   & \qquad           & 581  & 12.35  &  c-AGN \\
\qquad         & PLSuperExpCutoff & 3.5 $\pm$ 0.4    & 0.4 $\pm$ 0.2     & 2.2 $\pm$ 0.3    & 733  &        & N \\	
\hline
J1544.1$-$2555 & PowerLaw         & 7.9 $\pm$ 0.7    & 2.36 $\pm$ 0.06   & \qquad           & 265  & 6.17   &   \\
\qquad         & PLSuperExpCutoff & 6.1 $\pm$ 0.7    & 1.5 $\pm$ 0.2     & 2.2 $\pm$ 0.6    & 300  & \qquad & \\
\hline 
J1624.2$-$4041 & PowerLaw         & 27 $\pm$ 1       & 2.36 $\pm$ 0.03   & \qquad           & 1049 & 9.61   & c-MSP \\
\qquad         & PLSuperExpCutoff & 19.7 $\pm$ 0.9   & 1.58 $\pm$ 0.04   & 2.8 $\pm$ 0.1    & 1082 & \qquad & \\	
\hline
J1626.2$-$2428c& PowerLaw         & 26 $\pm$ 3       & 2.50 $\pm$ 0.05   & \qquad           & 587  & 3.17   & c-AGN \\
\qquad         & PLSuperExpCutoff & 23 $\pm$ 3       & 2.3 $\pm$ 0.1     & 10 $\pm$ 5       & 542  &        & N \\
\hline
J1645.7$-$2149 & PowerLaw         & 10 $\pm$ 1       & 2.54 $\pm$ 0.07   & \qquad           & 176  & 3.68   & \\
\qquad         & PLSuperExpCutoff & 9 $\pm$ 1        & 1.9 $\pm$ 0.3     & 2 $\pm$ 1        & 188  & \qquad & \\
\hline
J1649.6$-$3007 & PowerLaw         & 7.1  $\pm$ 0.8   & 2.24 $\pm$ 0.06   & \qquad           & 208  & 6.23   & \\
\qquad         & PLSuperExpCutoff & 4.2  $\pm$ 0.8   & 1.2 $\pm$ 0.3     & 2.6 $\pm$ 0.7    & 244  & \qquad & \\	
\hline
J1702.8$-$5656 & PowerLaw         & 34 $\pm$ 1       & 2.53 $\pm$ 0.03   & \qquad           & 2209 & 6.65   &   \\
\qquad         & PLSuperExpCutoff & 32 $\pm$ 1       & 2.21 $\pm$ 0.07   & 5 $\pm$ 1        & 2228 &        & N \\	
\hline 
J1744.1$-$7619 & PowerLaw         & 18.5 $\pm$ 0.7   & 2.15 $\pm$ 0.03   & \qquad           & 2023 & 14.72  & c-MSP  \\
\qquad         & PLSuperExpCutoff & 13.6 $\pm$ 0.8   & 1.1 $\pm$ 0.1     & 1.9 $\pm$ 0.2    & 2211 & \qquad &  \\
\hline
J1753.6$-$4447 & PowerLaw         & 7.5  $\pm$ 0.7   & 2.24 $\pm$ 0.06   & \qquad           & 250  & 6.10   &   \\
\qquad         & PLSuperExpCutoff & 5.0  $\pm$ 0.7   & 1.2 $\pm$ 0.3     & 2.1 $\pm$ 0.6    & 286  & \qquad & \\	
\hline
J1757.7$-$6030 & PowerLaw         & 2.6 $\pm$ 0.4    & 2.00 $\pm$ 0.08   & \qquad           & 160  & 5.14   & \\
\qquad         & PLSuperExpCutoff & 1.2 $\pm$ 0.3    & 0.8 $\pm$ 0.4     & 4 $\pm$ 1        & 187  & \qquad & \\
\hline
J1808.3$-$3357 & PowerLaw         & 10 $\pm$ 1       & 2.43 $\pm$ 0.06   & \qquad           & 186  & 7.39   & \\
\qquad         & PLSuperExpCutoff & 6.9 $\pm$ 0.8    & 1.1 $\pm$ 0.2     & 1.3 $\pm$ 0.2    & 237  & \qquad & \\
\hline	 
J1831.6$-$6503 & PowerLaw         & 3.2 $\pm$ 0.4    & 2.08 $\pm$ 0.07   & \qquad           & 177  & 6.12   & \\
\qquad         & PLSuperExpCutoff & 1.7 $\pm$ 0.4    & 0.6 $\pm$ 0.4     & 2.0 $\pm$ 0.7    & 213  & \qquad & \\
\hline
J1946.4$-$5403 & PowerLaw         & 8.9 $\pm$ 0.5    & 2.20 $\pm$ 0.04   & \qquad           & 795  & 11.49  & c-MSP  \\
\qquad         & PLSuperExpCutoff & 6.2 $\pm$ 0.5    & 0.8 $\pm$ 0.2     & 1.3 $\pm$ 0.2    & 927  & \qquad &   \\
\hline
J2039.6$-$5618 & PowerLaw         & 13.0 $\pm$ 0.6   & 2.17 $\pm$ 0.03   & \qquad           & 1517 & 8.59   & c-MSP \\
\qquad         & PLSuperExpCutoff & 10.6 $\pm$ 0.7   & 1.61 $\pm$ 0.09   & 4.4 $\pm$ 0.8    & 1571 & \qquad &  \\
\hline
J2043.8$-$4801 & PowerLaw         & 3.6 $\pm$ 0.5    & 2.20 $\pm$ 0.08   & \qquad           & 189  & 5.85   & \\
\qquad         & PLSuperExpCutoff & 2.0 $\pm$ 0.4    & 0.8 $\pm$ 0.4     & 1.8 $\pm$ 0.5    & 221  & \qquad & \\
\hline
J2112.5$-$3044 & PowerLaw         & 13.5 $\pm$ 0.6   & 2.01 $\pm$ 0.03   & \qquad           & 2053 & 13.54  & c-subhalo \\
\qquad         & PLSuperExpCutoff & 9.7 $\pm$ 0.6    & 1.1 $\pm$ 0.1     & 3.0 $\pm$ 0.4    & 2237 & \qquad &  \\
\hline
J2131.1$-$6625 & PowerLaw         & 5.1 $\pm$ 0.5    & 2.42 $\pm$ 0.08   & \qquad           & 243  & 4.12   & \\
\qquad         & PLSuperExpCutoff & 4.0 $\pm$ 0.6    & 1.6 $\pm$ 0.3     & 2.3 $\pm$ 0.8    & 258  & \qquad & \\
\hline
J2133.0$-$6433 & PowerLaw         & 5.4 $\pm$ 0.6    & 2.27 $\pm$ 0.07   & \qquad           & 283  & 6.31   & c-subhalo \\
\qquad         & PLSuperExpCutoff & 3.6 $\pm$ 0.5    & 1.1 $\pm$ 0.3     & 1.7 $\pm$ 0.5    & 311  & \qquad &  \\
\hline
\hline
\end{tabular}
\end{table}

\clearpage
\begin{table} 
\begin{minipage}[]{100mm}
\caption[]{Source without sufficient curvature significance.
\label{tab:nosc}}
\end{minipage}
\centering
\fontsize{9}{9}\selectfont
\begin{tabular}{llccccc}
\hline
Source name  & Spectra model    & Flux/$10^{-9}$                 & $\Gamma$       & E$_{c}$      & TS     & Signif\_Curve \\
\qquad       &\qquad   & (photon\,cm$^{-2}$\,s$^{-1}$) & \qquad         & (GeV)        & \qquad & ($\sigma$)    \\
\hline
J0737.2-3233 & PowerLaw         & 14 $\pm$ 1 & 2.53 $\pm$ 0.06     & \qquad          & 326  & 2.87   \\
\qquad       & PLSuperExpCutoff & 13 $\pm$ 1 & 2.1 $\pm$ 0.2       & 3 $\pm$ 1       & 330  & \qquad \\	
\hline
\end{tabular}
\end{table}

\clearpage
\begin{table} 
	\begin{minipage}[]{100mm}
		\caption[]{Sources without clean background. \label{tab:bgs}}
	\end{minipage}
	\fontsize{9}{9}\selectfont
	\centering
	\begin{tabular}{llcccccc}	
		\hline
		Source name       & Spectra model & Flux density/10$^{-12}$  & $\Gamma$ & E$_{0}$ & Signif\_Avg & Signif\_Curve & \qquad  \\
		\qquad        & \qquad & (photon\,cm$^{-2}$MeV$^{-1}$s$^{-1}$) & \qquad   & (MeV)   & ($\sigma$)  & ($\sigma$)    & \qquad \\
		\hline
		J0456.2$-$6924 & PowerLaw      & 0.88 $\pm$ 0.10         & 2.3   & 1057 & 9.9       & 3.6   & \qquad  \\	
		J0816.1$-$5044 & PowerLaw      & 2.40  $\pm$ 0.32         & 2.5   & 771  & 7.75       & 3.6   & \qquad  \\
		J0905.8$-$2127 & PowerLaw      & 4.93  $\pm$ 0.75         & 2.7   & 477  & 6.6       & 3.3   & \qquad  \\
		J1256.1$-$5703 & PowerLaw      & 5.16  $\pm$ 0.77         & 2.7   & 544  & 6.4       & 3.3   & \qquad  \\
		J1408.0$-$2924 & PowerLaw      & 6.46  $\pm$ 0.87         & 2.7   & 419  & 7.4       & 3.3   & \qquad  \\
		J1820.4$-$3217 & PowerLaw      & 1.00 $\pm$ 0.16         & 2.3   & 1011 & 6.2       & 3.3   & \qquad  \\
		\hline
		\hline	
		Source Name  & Spectra Model & Flux\_Density                 & $\alpha$ & $\beta$ & E$_{b}$ & Signif\_Avg & Signif\_Curve \\
		\qquad       & \qquad        & ($10^{-12}$photon cm$^{-2}$MeV$^{-1}$s$^{-1}$) & \qquad & \qquad & (MeV)  & \qquad   & \qquad \\
		\hline
		J1128.6$-$5434 & LogParabola   & 29.1  $\pm$ 3.4          & 2.56   & 0.94  & 336  & 8.5       &    4.2       \\
		J1557.0$-$4225 & LogParabola   & 35.3  $\pm$ 3.7          & 2.33   & 0.55  & 369  & 11.5      &    4.4       \\
		J1729.7$-$2408 & LogParabola   & 16.6  $\pm$ 1.7          & 2.47   & 0.40  & 567  & 11.7      &    4.4       \\
	\hline
\end{tabular}
\end{table}

\clearpage
\begin{table}
\begin{minipage}[]{100mm}
  \caption[]{Log and fitting results of X-ray sources in the fields of the candidate MSPs using a PL model.
  \label{xray}}
\end{minipage}
\tiny
\centering
  \begin{tabular}{llcccccccccc}
\hline
Source                  & Data             & ObsID           & Exp                  & X-ray    & R.A.           & Dec.                  & $N_{H}$                  & $\Gamma$ & $F_{\rm X}$ & $\chi^2/\rm dof$ & GXr ($G_{100}$) \\
            &  \qquad        & \qquad         & (ks)        & Source & (h:m:s)      & (${^\circ}:':''$)  & ($10^{21}$) &  \qquad       & ($10^{-14}$) & \qquad       &  \qquad  \\

\hline
J0802.3 & {\it XMM}     & 0691980301 & 18.1             & X1         & 8:02:16.56 & -56:11:51.00 & 1.50       & $1.7 (f)$                       & $0.66^{+0.95}_{-0.00}$ & 11.4/4(C) & 1333 (8.8) \\
\qquad	       & \qquad		& \qquad          & \qquad	     & X2         & 8:02:08.64 & -56:12:13.68 & 1.50       & $2.14^{+0.93}_{-0.74}$ & $2.62^{+1.47}_{-0.99}$ & 0.9/5 (C) & 336\\
\qquad	       & \qquad		& \qquad	  & \qquad	     & X3         & 8:02:13.92 & -56:13:21.36 &	1.50 & $1.93^{+0.83}_{-0.72}$ & $2.35^{+1.69}_{-0.99}$ & 2.3/4(C) &  374\\
\qquad	       & \qquad		& \qquad          & \qquad	     & X4         & 8:01:59.04 & -56:11:24.00 &	1.50 & $0.90^{+1.08}_{-0.98}$ & $3.18^{+3.93}_{-1.82}$ & 0.2/2(C) & 277\\
\qquad	       & \qquad		& \qquad	  & \qquad	     & X5         & 8:02:19.20 & -56:14:37.32 &	1.50 & $2.59^{+1.52}_{-0.95}$ & $2.86^{+3.32}_{-1.28}$ & 2.2/2(C) &  308\\
J0838.8 & {\it Swift} 	& 00041343002 & 4.2             & X1         & 8:38:43.20 & -28:27:01.45 & 1.39       & $1.70^{+0.18}_{-0.18}$ & $466.0^{+54.5}_{-51.4}$ & 15.1/14 &  2 (9.4) \\
J0933.9 & {\it Chandra}& 14813             & 43.6           & X1         & 9:34:00.58 & -62:33:52.44 & 2.04       & $4.05$                            & $-$                                   & 25.9/7* & (12) \\
J0954.8 & {\it Swift} 	& 00031664001 & 3.6           & X1            & 9:55:27.78	& -39:47:49.84 & 1.35 & $0.32^{+1.21}_{-1.35}$ & $108.0^{+410.0}_{-76.2}$ & 1.4/3(C) & 16 (17)\\
J1035.7 & {\it XMM} & 0692830201       & 24.9            & X1        & 10:35:27.60 & -67:20:15.36 & 1.80      & $2.86^{+0.76}_{-0.64}$ & $2.69^{+1.34}_{-0.92}$ & 11.7/14(C) & 781 (21) \\
J1119.9 & {\it XMM}     & 0742930101  & 73.6            & X1          & 11:19:58.08 & -22:04:57.00 & 0.37 & $2.21^{+0.15}_{-0.15}$ & $14.10^{+1.29}_{-1.27}$ & 45.8/33 & 128 (18) \\
\qquad	       & \qquad		 &\qquad	      & \qquad      & X2          & 11:20:01.68 & -22:04:55.92 & 0.37 & $1.70^{+0.31}_{-0.30}$ & $5.24^{+1.32}_{-1.10}$ & 3.4/8 &  343\\
\qquad	       & \qquad		 &\qquad	      & \qquad       & X3          & 11:19:59.04 & -22:03:15.48 & 0.37 & $1.12^{+0.62}_{-0.70}$ & $2.04^{+1.34}_{-0.82}$ & 3.3/3(C) & 1607\\
J1624.2 & {\it XMM}      & 0722940101 & 31.0           & X1           & 16:24:09.84 & -40:44:25.08 & 2.60 & $-0.27^{+0.62}_{-0.81}$ & $4.76^{+1.87}_{-1.58}$ & 4.8/7(C) &  567 (27) \\
\qquad	       & \qquad		&\qquad	     &	 \qquad               & X2           & 16:24:07.68 & -40:44:37.32 & 2.60 & $1.21^{+0.58}_{-0.56}$ & $2.22^{+1.05}_{-0.78}$ & 9.2/6(C) & 1216\\
\qquad	       & \qquad		&\qquad	     &	 \qquad              & X3            & 16:24:02.16 & -40:45:44.28 & 2.60	& $1.41^{+0.44}_{-0.49}$ & $2.56^{+0.94}_{-0.75}$ & 10.8/6(C) & 1055\\
\qquad	       & \qquad		&\qquad	     &	  \qquad             & X4            & 16:24:15.12 & -40:47:12.48 & 2.60	& $0.96^{+0.88}_{-0.92}$ & $1.92^{+1.35}_{-1.06}$ & 8.8/8(C) & 1406\\
\qquad	       & \qquad		&\qquad	     &	  \qquad             & X5            & 16:24:09.36 & -40:42:38.88 & 2.60	& $1.69^{+0.80}_{-0.79}$ & $0.98^{+0.65}_{-0.51}$ & 10.4/8(C) & 2755\\ 
\qquad	       & \qquad		&\qquad	     &	  \qquad             & X6            & 16:24:04.56 & -40:47:20.76 & 2.60	& $6.18^{+3.15}_{-2.80}$ & $34.40^{+716.0}_{-31.3}$ & 8.2/8(C) &  78\\
\qquad	       & \qquad		&\qquad	     &	  \qquad             & X7            & 16:24:12.48 & -40:47:57.12 & 2.60	& $0.97^{+0.56}_{-0.58}$ & $3.78^{+1.86}_{-1.42}$ & 4.9/8(C) &  714\\
\qquad	       & \qquad		&\qquad	     &	  \qquad             & X8            & 16:24:26.16 & -40:45:40.32 & 2.60	& $1.78^{+1.42}_{-1.14}$ & $1.16^{+0.89}_{-0.64}$ & 15.7/7(C) & 2328\\
J1626.2 & {\it Chandra} 	& 17249	    & 100.0       & X1            & 16:26:48.49 & -24:28:38.91 & 1.34 & $1.08^{+0.60}_{-0.55}$ & $3.43^{+0.86}_{-0.74}$ & 7.5/6(C) & 845 (29) \\
\qquad	        & \qquad	&\qquad	     &	 \qquad              & X2            & 16:26:40.49 & -24:27:15.13 & 1.34 & $0.56$                            & $-$                                  & 59.0/17* &  \\
\qquad	        & \qquad	& 637 	     & 97.7                & X1            & 16:26:48.41 & -24:28:36.93 & 1.34 & $0.68^{+0.75}_{-0.73}$ & $2.13^{+1.13}_{-0.82}$ & 9.6/6(C) & 1361\\
\qquad	        & \qquad	&\qquad	     &	 \qquad             & X2            & 16:26:40.48 & -24:27:14.60 & 1.34 & $0.63$                             & $-$                                  & 23.5/9* & \\ 
J1946.4 & {\it Suzaku} 	& 706026010 & 42.4             & X1            & 19:46:33.82 & -54:02:37.23 & 0.37 & $1.29^{+0.29}_{-0.28}$ & $12.7^{+2.36}_{-2.26}$ & 23.7/18 & 72 (9.2) \\
J2112.5 & {\it XMM} 	& 0672990201 & 33.8           & X1            & 21:12:32.16 & -30:44:04.92 & 0.66 & $2.20^{+0.73}_{-0.59}$ & $1.48^{+0.78}_{-0.54}$ & 7.7/5 & 1284 (19) \\
\hline
\hline
\end{tabular}
{Notes:* denotes the cases where the reduced $\chi^2$ $> 2$ and we tried other spectral models. (1) Source Name; (2) Data; (3) ID of the observations; 
(4) Exposure time in kilo-seconds for each observation; (5) X-ray sources in the $2\sigma$ error circle; (6)--(7) Right Ascension (R.A.; J2000.0) and Declination (Dec.; J2000.0); 
8) Galactic absorption column density in units of $\rm cm^{-2}$; (9) Photon index; (10) Unabsorbed flux in 0.3-10 keV band in units of $\rm erg~cm^{-2}~s^{-1}$; 
(11) $\chi^2/\rm dof$ value for the model, where C-statistics is indicated by C.; (12) $\gamma$-ray to X-ray flux ratios (GXr) for each candidate MSP with X-ray sources detected in the field, where the $\gamma$-ray 0.1--100 GeV flux ($G_{100}$) of each candidate MSP, in units of $10^{-12}\rm erg~cm^{-2}~s^{-1}$, is given in parentheses.}
\end{table}


\clearpage
\begin{table}
\begin{minipage}[]{100mm}
  \caption[]{X-ray flux upper limits for 14 candidate MSPs.
  \label{nodetect}}
\end{minipage}
\tiny
\centering
  \begin{tabular}{llccccccccc}
\hline
Src            & R.A.        & Dec.         & Data        & ObsID       & Exp   & CR       & $N_{H}$ & $F^{upper}_{X}$ & $G_{100}/10^{-12}$ &$G_{100}/F^{upper}_{X}$ \\
Name           & (deg)       & (deg)        & \qquad      & \qquad      & Time  & ($\times 10^{-3}$) & ($\times 10^{20}$) & ($\times 10^{-13}$)& ($\rm erg~cm^{-2}~s^{-1}$) &  \qquad \\
\hline
J0048.1$-$6343 & 12.21070306 & -63.76885597 & {\it Swift} & 00047132001 & 3099  & $< 3.27$ & 1.97 & $< 1.34$ & 1.7  & $>13$\\
J1231.6$-$5113 & 187.9228875 & -51.2445342  & {\it Swift} & 00041384001 & 3602  & $< 2.09$ & 13.90 & $< 1.16$ & 12   & $>103$\\
J1400.2$-$2413 & 210.0116122 & -24.2486119  & {\it Swift} & 00047217004 & 2186  & $< 7.56$ & 5.33 & $< 3.51$ &  3.4  & $>9.7$\\ 
J1539.2$-$3324 & 234.8356327 & -33.41744465 & {\it Swift} & 00048054023 & 15084 & $< 0.81$ & 9.06 & $< 0.41$ &  1.0  & $>24$\\   
J1544.1$-$2555 & 236.0585575 & -25.9120224  & {\it Swift} & 00085021004 & 993   & $< 11.80$ & 11.00 & $< 6.24$ &  7.9  & $>13$\\
J1649.6$-$3007 & 252.4595489 & -30.1831453  & {\it Swift} & 00085034001 & 4155  & $< 5.35$ & 15.90 & $< 3.08$ &  7.1  & $>23$\\    
J1702.8$-$5656 & 255.6861421 & -56.91435142 & {\it Swift} & 00041424001 & 1214  & $< 6.39$ & 11.30 & $< 3.40$ & 37  & $>108$\\    
J1753.6$-$4447 & 268.3488964 & -44.76971147 & {\it Swift} & 00047264001 & 1567  & $< 8.63$ & 13.20 & $< 4.75$ &  7.6  & $>16$\\
J1757.7$-$6030 & 269.4143008 & -60.55432813 & {\it Swift} & 00047265005 & 1994  & $< 4.97$ & 7.19 & $< 2.42$ & 3.7  & $>15$\\    
J1808.3$-$3357 & 272.1154608 & -33.86901995 & {\it Swift} & 00047271002 & 7123  & $< 3.09$ & 18.10 & $< 1.84$ &  8.7  & $>47$\\    
J1831.6$-$6503 & 277.7440476 & -65.06402308 & {\it Swift} & 00047281002 & 1704  & $< 7.50$ & 6.70 & $< 3.61$ &  3.7   & $>10$\\   
J2043.8$-$4801 & 310.9822976 & -48.02857435 & {\it Swift} & 00047307004 & 1116  & $< 10.20$ & 2.83 & $< 4.35$ & 3.6  & $>8$\\   
J2131.1$-$6625 & 322.7121318 & -66.40198989 & {\it Swift} & 00085130006 & 845   & $< 9.51$ & 2.44 & $< 3.99$ &  4.9  & $>12$\\
J2133.0$-$6433 & 323.3171713 & -64.53428129 & {\it Swift} & 00047316004 & 3357  & $< 4.98$ & 3.09 & $< 2.15$ &  5.1  & $>23$\\
\hline
\end{tabular}
{Notes:(1) Source Name; (2)-(3) Right Ascension (R.A.; J2000.0) and Declination (Dec.; J2000.0) of each source;
(4) Data; (5) ID of Observation used in the analysis; (6) Exposure time in seconds for each observation;
(7) $3\sigma$ upper limit of count rate in unit of $\rm counts~s^{-1}$ (8) Galactic absorption column density in $\rm cm^{-2}$; 
(9) $3\sigma$ upper limit of flux in 0.3-10 keV band in units of $\rm erg~cm^{-2}~s^{-1}$; (10) $\gamma$-ray 0.1--100 GeV flux; (11) Lower limit on the $\gamma$-ray to X-ray flux ratio.}
\end{table}

\clearpage

\end{document}